\documentclass[jkps,preprint,fleqn,showkeys]{revtex4}
\usepackage{graphicx}%
\usepackage{multirow}%
\usepackage{amsmath,amssymb,amsfonts}%
\usepackage{amsthm}%
\usepackage{mathrsfs}%
\usepackage[title]{appendix}%
\usepackage{xcolor}%
\usepackage{textcomp}%
\usepackage{booktabs}%
\usepackage{algorithm}%
\usepackage{algorithmicx}%
\usepackage{algpseudocode}%
\usepackage{listings}%
\usepackage{physics}%
\usepackage{tikz, orcidlink}%
\begin{document}
\setcounter{page}{0}
\title[]{Exciton Energy Transfer in Conjugated Polymers via a Three-Level Donor–Acceptor Model}

\author{Hyeonseok \surname{Yang}\orcidlink{0009-0007-1308-4724}}
\email{yhs1802@gmail.com}
\affiliation{Department of Chemistry, Graduate School of Science, Kyoto University, Kyoto, 606-8502, Japan}

\date{\today}

\begin{abstract}
We investigate intra-chain exciton energy transfer (EET) in a conjugated polymer using a reduced three-level donor–acceptor model that separates vertical excitations from vibrationally relaxed states. Population dynamics are obtained with the hierarchical equations of motion (HEOM), enabling a non‑Markovian treatment of system–bath (S--B) interactions. Case studies on donor--acceptor (D--A) pairs with small and large energy gaps reveal competition among coherent hopping, vibrational relaxation, and Förster resonance energy transfer (FRET). Varying the hopping and the S--B relaxation rate highlights roles of exciton transfer, thermal relaxation, and dipole‑mediated coupling. The results show that energy gap, hopping, and relaxation jointly control intrachain EET.
\end{abstract}


\maketitle

\section{Introduction}\label{sec:intro}

MEH-PPV (poly[2-methoxy-5-(2-ethylhexyloxy)-1,4-phenylenevinylene]) is a conjugated polymer that has been widely studied as a reference molecular system for excited-state photophysics. \cite{burroughes1990light,huser2000single,panzer2017temperature} One of the characteristic features of this system is that the lineshape of the photoluminescence spectrum does not directly mirror the absorption spectrum. The emission occurs within a specific energy range, nearly independent of the absorbed energy.
This observation implies that exciton populations undergo energy redistribution such that only exciton states within a specific energy band contribute to photoluminescence. \cite{okazaki2024generation,okazaki2023luminescence,kim2006detailed,yu2004low}

The spectral features have been interpreted in terms of inter-chain interactions in conjugated polymers, often within the framework of HJ-aggregate models.\cite{huser2000single,spano2010spectral,yamagata2012interplay,yamagata2012designing,hestand2015interference,marletta2004photoluminescence,chakraborty2016role,yamagata2013red,milward2018extracting} However, it is well understood that intra-chain interactions play a crucial role. Specifically, torsional disorder along the backbone locally breaks $\pi-\pi$ conjugation and fragments the chain into independently conjugated segments known as chromophores. \cite{barford2022exciton,dimitriev2022dynamics,schindler2004universal} Exciton energy transfer (EET) is experimentally known to take place between such chromophores. We therefore focus on intra-chain EET and adopt the minimal unit of Donor--Acceptor (D--A) model; $i.e.$ $D^* + A \to D + A^*$ as a representative pair of two such chromophore segments.

From a theoretical perspective, recent advances in the study of electron and exciton transfer involving multiple electronic states have emphasized the need for nonperturbative, time-resolved approaches in condensed-phase systems.
\cite{tanaka2009quantum,tanaka2010multistate,cainelli2021exciton} In particular, when excitonic states contain both Frenkel-like and charge-transfer-like contributions, the interplay between electronic couplings and environmental fluctuations governs population relaxation, coherence decay, and transient emission responses.
\cite{sakamoto2017exciton,cainelli2022effect}

To account for these processes, we employ the hierarchical equations of motion (HEOM), which provides a non-Markovian description of open quantum systems. Using this method, the time evolution of the reduced density matrix can be treated explicitly, enabling us to describe dynamics of the system under general S--B interactions, as well as in the presence of external electromagnetic or gauge fields. \cite{yang2025theory,yang2025quantum,tanimura2012reduced,tanimura2014reduced,ishizaki2009unified,ishizaki2009adequacy,tanimura2020numerically}

The paper is organized as follows. Section~2 outlines the three-level D--A model and the HEOM-based theoretical framework. Section~3 describes the parameterization of the model for MEH-PPV chromophores and the numerical computation. Section~4 presents numerical results for representative cases, highlighting the roles of hopping strength, energy gap, and S--B coupling. Section~5 concludes with a summary of the main results.

\section{Theory}
\label{sec:theory}

We describe the theoretical framework for modeling intra-chain energy transfer in conjugated polymers.
Our approach is based on a reduced D--A description,
where the system is represented by a Frenkel--Holstein Hamiltonian that captures both short-range hopping and long-range Coulombic interactions among chromophores.
The interaction with the environment is incorporated through an S--B coupling and treated using the HEOM, enabling a non-Markovian description of the population dynamics.

\subsection{Model Hamiltonian}

\begin{figure}
    \centering
    \includegraphics[width=0.8\linewidth]{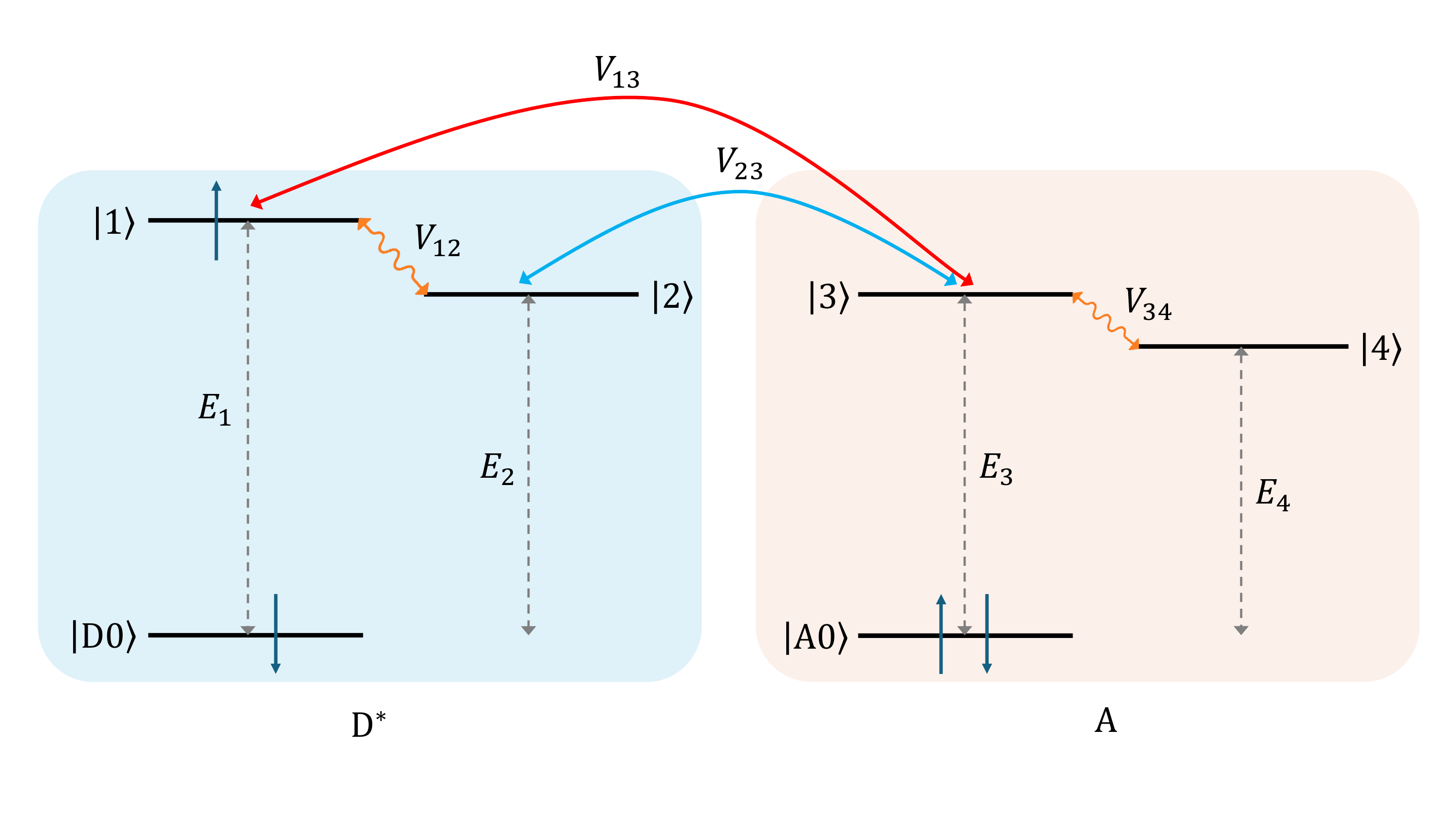}
    \caption{Schematic illustration of exciton energy transfer from a donor to an acceptor. After the vertical excitation from the donor ground state $\ket{\mathrm{D}0}$, the system is initialized in $\ket{1}$. Here, $V_{13}$ denotes the hopping constant for exciton transfer from the donor to the acceptor, while $V_{12}$ and $V_{34}$ represent intramolecular vibrational relaxation between states $\ket{1}$ and $\ket{2}$, and between $\ket{3}$ and $\ket{4}$, respectively. The coupling $V_{23}$ corresponds to the transfer constant from donor $\ket{2}$ to acceptor $\ket{3}$ mediated by transition dipole–dipole interactions.}
    \label{fig:DnAThreeLevel}
\end{figure}

We employ a three-level D–A model that includes vertical excitation from the chromophore's ground states (denoted as $\ket{\rm D0}$ and $\ket{\rm A0}$) to its vertical excited states (denoted as $\ket{1}$ and $\ket{3}$), along with vibrationally relaxed states (denoted as $\ket{2}$ and $\ket{4}$). Each basis state $\ket{\mathrm{n}}$ is a composite of an electronic state and a nuclear configuration, so that the primary intramolecular vibrational mode is internalized in a system part.
For the initial condition, we assume that the donor chromophore absorbs the applied light energy, leading to an exciton population of 1 in the excited state $\ket{1}$. (See Figure~\ref{fig:DnAThreeLevel}) 
Exciton transfer between $\ket{1}$ and $\ket{3}$ corresponds to a hopping interaction, while transitions from $\ket{1}$ to $\ket{2}$ and from $\ket{3}$ to $\ket{4}$ represent vibrational relaxation of the excited states. Additionally, we consider Förster resonance energy transfer (FRET) between $\ket{2}$ (as an emission state) and $\ket{3}$ (as an absorption state), which is relatively slow rather than the hopping interaction.

We write the total system Hamiltonian as
\begin{align}
H = H_S + H_B + H_I,
\end{align}
where $H_S$ describes the excitonic degrees of freedom of the D--A system, $H_B$ represents the vibrational environment (heat bath), and $H_I$ accounts for the S--B interaction.

The system Hamiltonian is given by
\begin{align}
H_S = \sum_{n} E_{n}\ketbra{n}{n}
+ \sum_{n\neq m}V_{nm}\left(\ketbra{m}{n} + \ketbra{n}{m}\right),
\end{align}
where $\{\ket{n}\}$ denote the excitonic basis states of the model.
Here, $E_n$ is the site energy of the excited state $\ket{n}$.
The reduced Planck constant is defined as $\hbar = h/2\pi$, where $h$ is Planck's constant.
The interaction matrix $V_{nm}$ denotes the excitonic transfer or vibrational relaxation coupling between states $\ket{n}$ and $\ket{m}$.

The bath Hamiltonian and the S--B interaction are written as \cite{seibt2021strong}
\begin{align}
H_B + H_I = \sum_{j}\hbar\omega_j\hat{b}^\dagger_j\hat{b}_j
+ \sum_{n}\sum_{j}\hbar\omega_j\sqrt{S_n} \Big(\hat{b}^\dagger_j+\hat{b}_j\Big)\ketbra{n}{n}.
\end{align}
Here, $\hbar\omega_j$ denotes the energy of a harmonic oscillator with mode $j$, which is associated with molecular vibrational modes and relevant to the thermal environment.
The operator $\hat b_j^{(\dagger)}$ is the annihilation (creation) operator of the $j$th harmonic oscillator.
The dimensionless Huang–Rhys factor $S_n$ was assumed to be identical to $S_0$ for all sites, since each chromophore is part of the same polymer chain and consists of identical monomers.

Because FRET involves the Coulombic coupling between the emission transition dipole of the donor and the absorption transition dipole of the acceptor, the dipole--dipole coupling between $\ket{2}$ and $\ket{3}$ is given by
\begin{align}
V_{23} = \frac{1}{4\pi\epsilon}\frac{\vec\mu_2 \cdot \vec\mu_3
- 3(\vec\mu_2\cdot\hat r_{DA})(\vec\mu_3\cdot\hat r_{DA})}{r_{DA}^3},
\end{align}
where $\vec\mu_2$ and $\vec\mu_3$ denote the transition dipole moment of states $\ket{2}$ and $\ket{3}$, respectively.
$r_{DA}$ is the distance between donor and acceptor dipoles, and $\hat r_{DA}$ is a unit vector that represents the direction between D--A. \cite{forster1948zwischenmolekulare}
The dielectric permittivity $\epsilon$ is given by the product of the vacuum permittivity $\epsilon_0$ and the relative dielectric constant $\epsilon_r$.

\subsection{HEOM}

We describe the exciton transfer dynamics by propagating the reduced density
operator under the total Hamiltonian using the hierarchical equations of motion
(HEOM).\cite{tanaka2009quantum,tanaka2010multistate,cainelli2021exciton,sakamoto2017exciton,cainelli2022effect,
shi2009electron,ishizaki2005quantum}
Since the system operates at finite temperature and is coupled to a heat bath,
HEOM provides a quantitatively reliable and nonperturbative framework for
capturing the dynamics.

The interaction with bath is described by the spectral distribution function (SDF);
\begin{equation}
    J_i(\omega) \equiv \sum_j c_{ij}^{2}\,\delta(\omega - \omega_j),
\end{equation}
where $c_{ij}$ denotes the coupling strength between the exciton level
at site $i$ and the $j$-th environmental mode of frequency $\omega_j$.
In our model, the intramolecular vibrational modes are treated explicitly as part of the system, while the low-frequency environmental fluctuations are described as an overdamped thermal bath~\cite{Bennett2018}. The dominant intramolecular vibrational mode, which underlies the vibronic progression, is incorporated at the level of the system states, so that the vertically excited and vibrationally relaxed configurations of each chromophore are included as separate basis states. Hence, including the ground state, this is termed the reduced three-level Donor-Acceptor model. Another mode associated with low-frequency environmental fluctuations, which modulate each electronic energy level, is treated independently as a thermal bath. Since these fluctuations are overdamped rather than sharply structured, they are well captured by the Drude--Lorentz form,
\begin{equation}
    J_i(\omega) = \frac{2\lambda_0\,\gamma_i\,\omega}{\omega^{2} + \gamma_i^{2}},
\end{equation}
where $\lambda_0$ denotes the reorganization energy and $\gamma_i$ denotes the Drude cutoff (S--B relaxation rate) associated with the site $i\in\{1,2,3,4\}$.

The bath correlation function is defined in terms of the spectral density as
\begin{align}
C_i(t)
=\frac{2}{\pi}\int_0^\infty d\omega\,J_i(\omega)
\left[\cos(\omega t)\coth\!\left(\frac{\beta\hbar\omega}{2}\right)
-i\sin(\omega t)\right].
\end{align}
To solve $C_i(t)$ in HEOM, we adopt the Padé spectrum decomposition (PSD) to obtain a compact exponential expansion of the bath correlation function, which enables stable propagation of non-Markovian system--bath dynamics. \cite{hu2010communication,zhu2011hierarchical,tian2010biexponential,hu2011pade}

The resulting HEOM for the auxiliary density operators 
$\rho^{(\mathbf{n})}$ takes the form
\begin{multline}
\frac{d}{dt}\rho^{(\mathbf{n})}
=-\left(\frac{i}{\hbar}\mathcal{L}_S
+\sum_{i,k}n_{ik}\nu_{ik}\right)\rho^{(\mathbf{n})} \\
-\frac{i}{\hbar}\sum_{i,k}\left[\Phi_i,
\rho^{(\mathbf{n+e}_{ik})}\right]
-\frac{i}{\hbar}\sum_{i,k}n_{ik}
\left(c_{ik}\Phi_i\rho^{(\mathbf{n-e}_{ik})}
-c_{ik}^*\rho^{(\mathbf{n-e}_{ik})}\Phi_i\right),
\end{multline}
where the system Liouvillian is defined as
$\mathcal{L}_S \rho = [H_S,\rho]$,
$\mathbf{n} \equiv \{ n_{ik} \}$ is a multi-index labeling the hierarchy with the truncation condition
$\sum_{i,k} n_{ik} \le N_{\mathrm{trun}}$,
and $\mathbf{e}_{ik}$ denotes the unit vector in the $(i,k)$ direction, where $k$ indicates the index of the Padé spectrum decomposition.
The operators $\Phi_i = \ketbra{i}{i}$ describe the system degrees of freedom coupled to the bath.
The coefficients $\nu_{ik}$ and $c_{ik}$ originate from the decomposition of the bath correlation function within the Padé spectrum decomposition.
In particular, $\nu_{i0} = \gamma_i$ corresponds to the Drude cut-off characterizing the S--B relaxation rate,
while $\{ \nu_{ik}, c_{ik} \}_{k>0}$ represent the characteristic frequencies and weights associated with PSD.

\section{Methods}
\label{sec:Meth} 

\begin{table}[t]
\centering
\renewcommand{\arraystretch}{1.0}

\begin{tabular}{c c c c c c c c c}
\hline\hline
$\mathrm{case}$ & $E_1$ & $E_2$ & $E_3$ & $E_4$ & $\lambda_0$ & $\hbar\gamma$ & $V_{13}$ \\
\hline
$5$--$7$
& 1.996 & 1.660 & 1.836 & 1.521
& 0.120 & \begin{tabular}{c}
Strong: 0.01\\
Weak: 0.001
\end{tabular}
& \begin{tabular}{c}
Strong: 0.7\\
Intermediate: 0.07\\
Weak: 0.007
\end{tabular}
\\
\hline
$2$--$10$
& 2.856 & 2.439 & 1.739 & 1.466
& 0.120 & \begin{tabular}{c}
Strong: 0.01\\
Weak: 0.001
\end{tabular}
& \begin{tabular}{c}
Strong: 0.7\\
Intermediate: 0.07\\
Weak: 0.007
\end{tabular}
\\
\hline\hline
\end{tabular}%

\caption{Model parameters used for the population-dynamics calculations.
$E_1$ and $E_3$ are vertical excitation energies, and $E_2$ and $E_4$ are vibrationally relaxed state energies for the donor and acceptor, respectively. These values are obtained from Table S1.
$\lambda_0$ is the reorganization energy.
$\gamma$ denotes the S--B relaxation rate (Drude cutoff) in the Drude--Lorentz spectral density.
All energy quantities ($E_i$, $\lambda_0$, $\hbar\gamma$ ,and $V_{13}$) are given in eV.}
\label{tab:model_parameters}
\end{table}

This study investigates the EET between two chromophores consisting of segments having two different conjugated lengths. To achieve this, we calculated electronic excitation energies for different chromophores, whose conjugated length varies by the number $n$ of repeat units.  
The ground and excited state geometries of 2- to 10-conjugated segments were fully optimized at the B3LYP/ 6-31+G(d,p) level. All calculations were performed using Gaussian 16 in the toluene solvent, employing the PCM implicit solvent method. \cite{g16} The results are summarized in Table~S1 and Fig. S1 and show a good agreement with previously reported studies, consistent with the finding that longer conjugation lengths lead to lower excitation energies in conjugated polymers. \cite{rissler2004effective}

In this study, we chose two representative cases to understand the mechanism of the EET. The case of $n = 5$ for the donor chromophore and $n = 7$ for the acceptor chromophore (5--7) represents the combination with a small energy gap. It is also reported that MEH-PPV polymer has an effective conjugation length of $n = 5$ at room temperature.\cite{kohler2012order} To compare a combination with a larger energy gap, we further chose $n = 2$ for the donor chromophore and $n = 10$ for the acceptor chromophore (2--10).
For each chromophore, only the lowest singlet vertical excitation $S_1$ is retained. In $D_{2h}$-symmetric PPV oligomers, $S_1$ excitation in an absorption process is dominant while higher transition states are forbidden or weakly allowed \cite{cornil1997photoluminescence,panda2013electronically}. Furthermore, any population reaching higher states decays rapidly to $S_1$ via internal conversion (Kasha's rule) before any inter-chromophore transfer occurs. Hence, higher excited states are therefore neglected.

The site energies of the donor and acceptor states were listed in Table~\ref{tab:model_parameters}.
The Huang--Rhys factor $S_0$ was set to $0.72$, as reported for MEH-PPV in toluene($\epsilon_r = 2.4$) at $300~\mathrm{K}$.\cite{yamagata2013red,quan2006solvent,cornil1997photoluminescence}
The dominant intramolecular vibrational mode was assigned a characteristic
frequency $\hbar\Omega = 0.173~\mathrm{eV}$ dominated by the C=C stretching mode. \cite{spano2005modeling, spano2016photophysical, janke2020frenkel} For the reorganization energy, we adopted the same order as the intramolecular vibrational reorganization energy,
$\lambda_0 \approx S_0\hbar\Omega$, merely to fix the
overall energy scale of the dissipative coupling.
The transfer constants $V_{nm}$ are inferred from experimental relaxation timescales via the order-of-magnitude relation $\Delta E \sim h/\Delta t$. From the $\sim 100$ fs vibrational-relaxation component reported for MEH-PPV \cite{dykstra2009conformational,sperling2008exciton,consani2015relaxation}, we set $V_{12} = V_{34} = 0.007~\mathrm{eV}$. For the inter-chromophore hopping $V_{13}$, the $\sim 30~$fs anisotropic transition timescale reported in Ref.\cite{sperling2008exciton} is taken as the physically realistic order, which maps the intermediate regime ($\sim10$ fs) to $V_{13} = 0.07~\mathrm{eV}$; the strong ($0.7~\mathrm{eV}$) and weak ($0.007~\mathrm{eV}$) regimes are then defined one order of magnitude above and below as limiting cases that bracket the physical range.
The dipole–dipole coupling $V_{23}$, computed with Gaussian 16, is $0.0035~\mathrm{eV}$ for 5–7 and $0.0024~\mathrm{eV}$ for 2–10. For each hopping regime, $\hbar\gamma$ was set to $0.01~\mathrm{eV}$ and $0.001~\mathrm{eV}$.

The numerical calculations of the population were implemented in Python using NumPy 1.26.3 and Numba 0.60.0.\cite{harris2020array,lam2015numba}
Time-dependent HEOM equations were solved via the fourth-order Runge--Kutta method.
All computations were executed on a system equipped with an Intel Core i9-13900H CPU.

\begin{figure}
    \centering
    \includegraphics[width=0.75\linewidth]{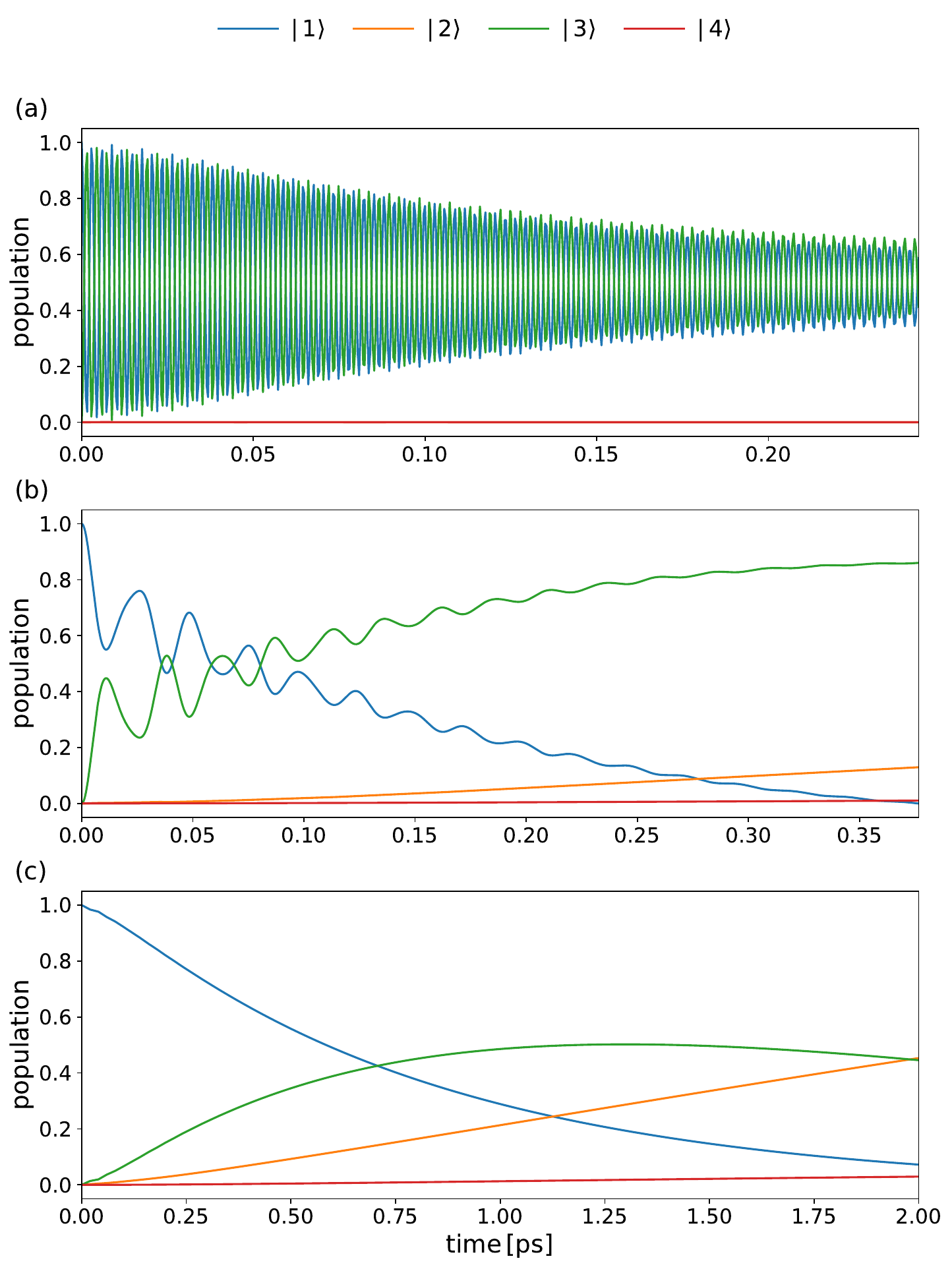}
    \caption{Time evolution of state populations for the 5--7 system under strong S--B relaxation rate ($\hbar\gamma = 0.01~\mathrm{eV}$). Panels (a)--(c) show the population dynamics calculated with different hopping constants: (a) $V_{13} = 0.7~\mathrm{eV}$, (b) $V_{13} = 0.07~\mathrm{eV}$, (c) $V_{13} = 0.007~\mathrm{eV}$. The system is initially prepared with the exciton fully localized in the donor($n=5$) state of $\ket{1}$, and the subsequent population transfer among the donor and acceptor excited and relaxed states ($\ket{1}$--$\ket{4}$) is shown as a function of time. Blue, orange, green, and red curves correspond to the populations of states $\ket{1}$, $\ket{2}$, $\ket{3}$, and $\ket{4}$, respectively.}

    \label{fig:5-7_0.01}
\end{figure}

\begin{figure}
    \centering
    \includegraphics[width=0.75\linewidth]{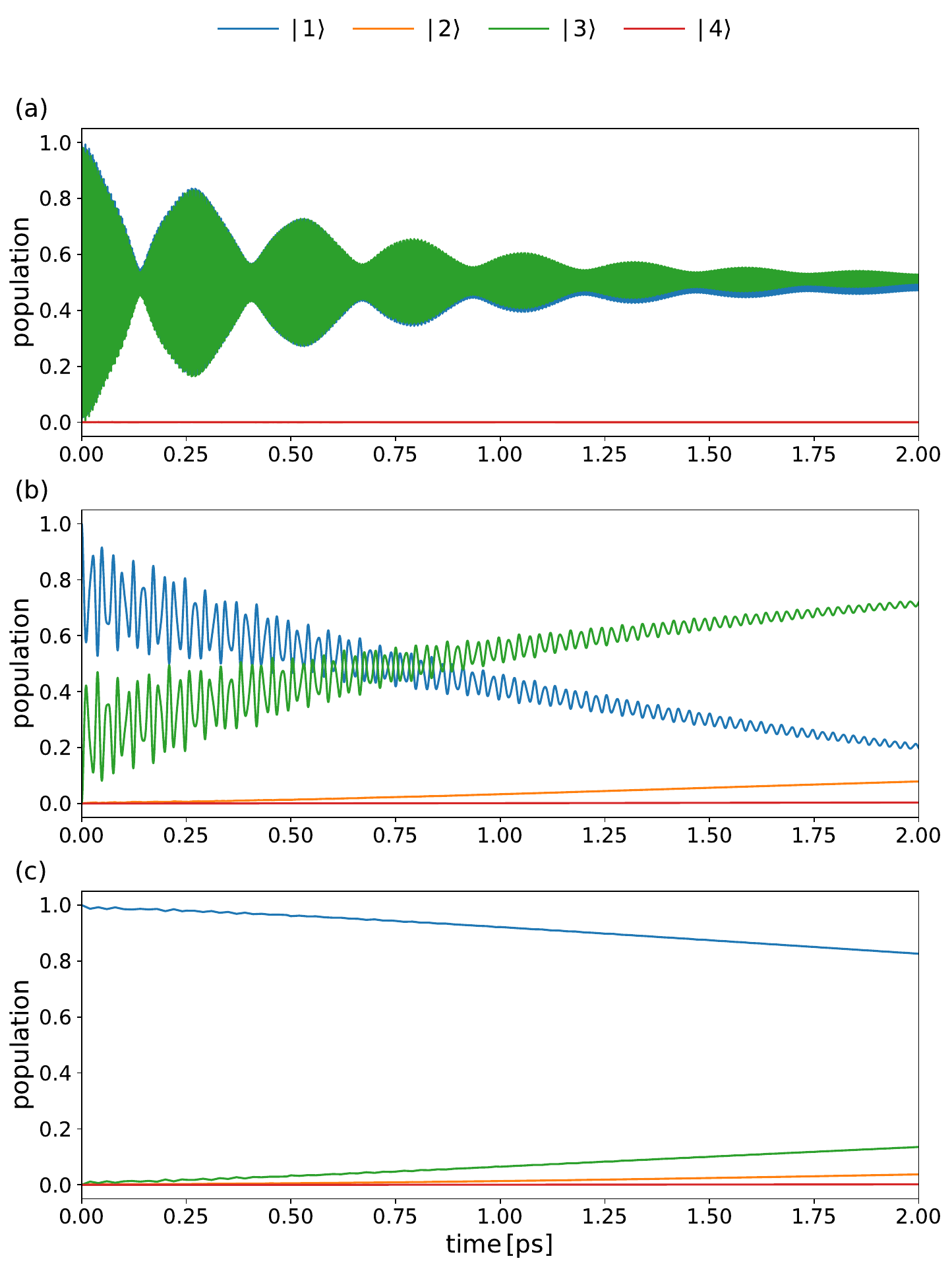}
    \caption{Time evolution of state populations for the 5--7 system under weak S--B relaxation rate ($\hbar\gamma = 0.001~\mathrm{eV}$). Each panel corresponds to (a) $V_{13} = 0.7~\mathrm{eV}$, (b) $V_{13} = 0.07~\mathrm{eV}$, (c) $V_{13} = 0.007~\mathrm{eV}$. All other conditions are the same as in Fig.~\ref{fig:5-7_0.01}.}
    \label{fig:5-7_0.001}
\end{figure}

\section{Results}

\subsection{The case of 5--7}

Figures~\ref{fig:5-7_0.01} and \ref{fig:5-7_0.001} show population dynamics for the $5$--$7$ system. 
In each figure, panels (a)--(c) correspond to strong, intermediate, and weak hopping, respectively; the S--B relaxation rate is fixed at $\hbar\gamma=0.01~\mathrm{eV}$ in Fig.~\ref{fig:5-7_0.01} and $\hbar\gamma=0.001~\mathrm{eV}$ in Fig.~\ref{fig:5-7_0.001}. 
In all cases, the system is initially prepared with the exciton fully localized in $\ket{1}$.

Figure~\ref{fig:5-7_0.01} represents the results under a large $\gamma$, where the dependence on the hopping is as follows.
In Fig.~\ref{fig:5-7_0.01}~(a), strong $V_{13}$ yields coherent population exchange between $\ket{1}$ and $\ket{3}$. With other pathways playing only a minor role, oscillatory populations dominate. 
In Fig.~\ref{fig:5-7_0.01}~(b), corresponding to a intermediate $V_{13}$, oscillations between $\ket{1}$ and $\ket{3}$ are still preserved at early times.
However, as time evolves, decoherence and energy relaxation associated with thermal dissipation progressively become dominant.
After the populations of $\ket{1}$ and $\ket{3}$ intersect, the dynamics change to a monotonic transfer regime.

Furthermore, in Fig.~\ref{fig:5-7_0.01}~(c), where $V_{13}$ becomes comparable to the S--B relaxation rate, coherent transfer is strongly suppressed, and the population dynamics are governed by the processes of localization.
As a result, the population gradually accumulates in the vibrationally relaxed states $\ket{2}$ and $\ket{4}$, while a FRET-mediated transfer between $\ket{2}$ and $\ket{3}$ provides a slow and secondary redistribution channel toward the acceptor.

The influence of $\gamma$ can be examined by comparing Figs.~\ref{fig:5-7_0.01} and \ref{fig:5-7_0.001}.
As shown in Fig.~\ref{fig:5-7_0.001}~(a), the thermal dissipation is weakened, allowing oscillations between $\ket{1}$ and $\ket{3}$  to persist over a longer time scale compared to Fig.~\ref{fig:5-7_0.01}~(a).
Similarly, in Fig.~\ref{fig:5-7_0.001}~(b), coherence appearing at early times remains longer than in Fig.~\ref{fig:5-7_0.01}~(b). It indicates that a small $\gamma$ weakens the process of vibrational relaxation on the population dynamics and consequently slows the localization.

\begin{figure}
    \centering
    \includegraphics[width=0.75\linewidth]{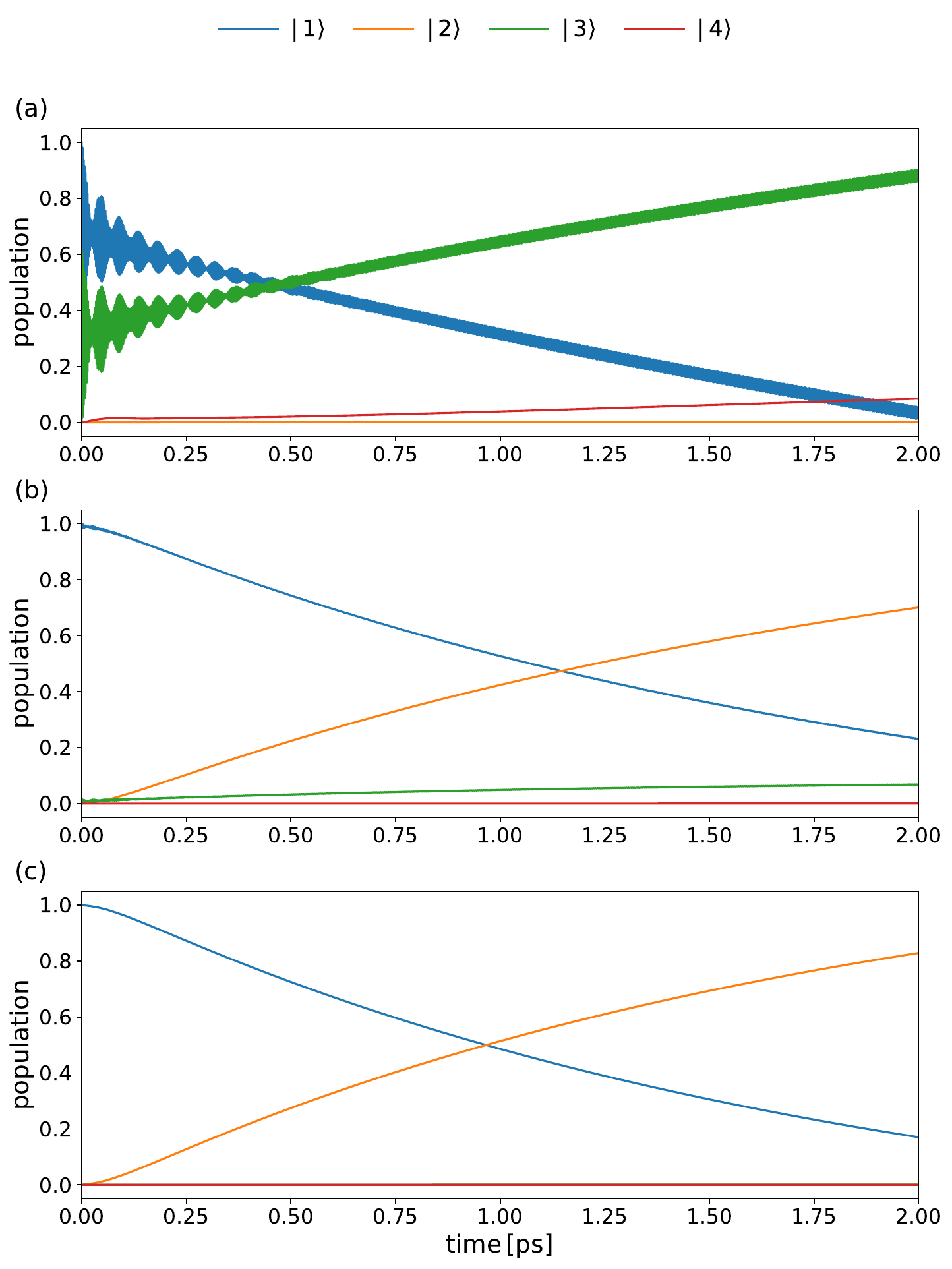}
    \caption{Time evolution of state populations for the 2--10 system under large S--B relaxation rate ($\hbar\gamma = 0.01~\mathrm{eV}$). Panels (a)--(c) show the population dynamics calculated with different hopping constants: (a) $V_{13} = 0.7~\mathrm{eV}$, (b) $V_{13} = 0.07~\mathrm{eV}$, (c) $V_{13} = 0.007~\mathrm{eV}$. The system is initially prepared with the exciton fully localized in the donor($n=2$) state of $\ket{1}$, and the subsequent population transfer among the donor and acceptor excited and relaxed states ($\ket{1}$--$\ket{4}$) is shown as a function of time. Blue, orange, green, and red curves correspond to the populations of states $\ket{1}$, $\ket{2}$, $\ket{3}$, and $\ket{4}$, respectively.}
    \label{fig:2-10_0.01}
\end{figure}

\begin{figure}
    \centering
    \includegraphics[width=0.75\linewidth]{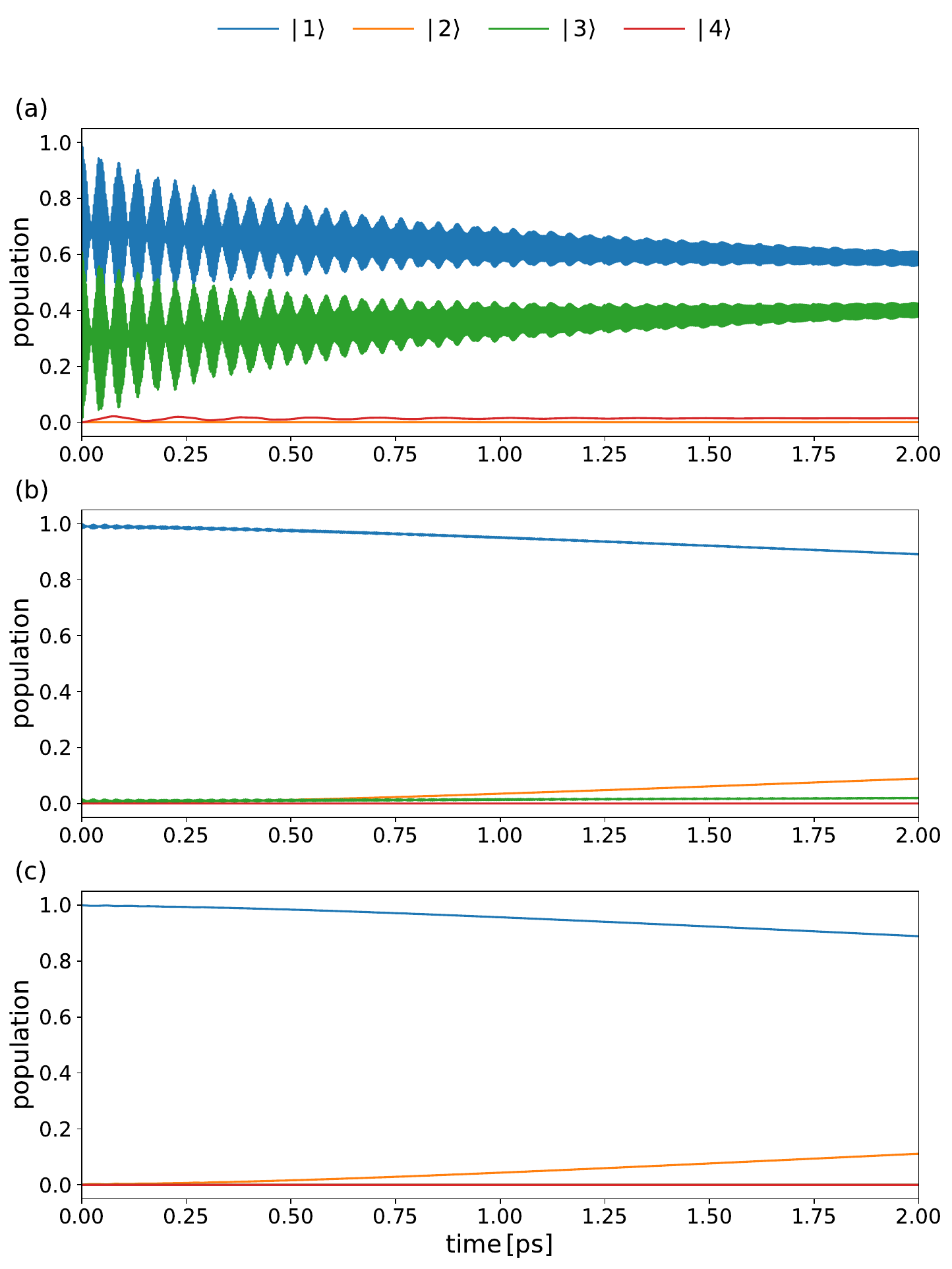}
    \caption{Time evolution of state populations for the 2--10 system under weak S--B relaxation rate ($\hbar\gamma = 0.001~\mathrm{eV}$). Each panel corresponds to (a) $V_{13} = 0.7~\mathrm{eV}$, (b) $V_{13} = 0.07~\mathrm{eV}$, (c) $V_{13} = 0.007~\mathrm{eV}$. All other conditions are the same as in Fig.~\ref{fig:2-10_0.01}.}
    \label{fig:2-10_0.001}
\end{figure}

\subsection{The case of 2--10}

Compared with the $5$--$7$ case, the $2$--$10$ combination lies in a regime with a large D--A energy difference, which is comparable to the strong-hopping coupling.
 
In Fig.~\ref{fig:2-10_0.01}~(a), despite the large energy gap, coherent oscillations between $\ket{1}$ and $\ket{3}$ are observed, which are gradually damped by $\gamma$. This indicates that sufficiently strong $V_{13}$ coupling can partially overcome the large energy gap between chromophores, resulting in early-time coherent transfer.
In Fig.~\ref{fig:2-10_0.01}~(b), oscillatory features are noticeably suppressed, and the dynamics become monotonic. Vibrational relaxation within the excited states occurs rather than exciton transfer between the two chromophores.
In Fig.~\ref{fig:2-10_0.01}~(c), hopping is further suppressed: the donor population remains thermally localized, the transfer between donor and acceptor is significantly reduced, and vibrational relaxation governs the rapid localization rather than sustained D--A exchange.

The influence of the S--B relaxation rate is assessed by comparing Fig.~\ref{fig:2-10_0.01} with Fig.~\ref{fig:2-10_0.001} (smaller $\gamma$). 
As shown in Fig.~\ref{fig:2-10_0.001}~(a), reduced decoherence allows oscillations between $\ket{1}$ and $\ket{3}$ to persist longer in the strong-hopping regime. 
However, for intermediate and weak hopping (Figs.~\ref{fig:2-10_0.001}~(b),(c)), populations remain dominated by vibrational relaxation, indicating that lowering $\gamma$ alone is insufficient to induce efficient D--A transfer when $V_{13}$ is too small.

\subsection{Discussion}
\label{Result.C}

The population dynamics presented above can still be compared with experimental observables at the level of timescales. First, under the intermediate-hopping condition for the 5--7 system (Fig.~2(b)), the donor population $|1\rangle$ decreases to below half within approximately 100--200 fs. This is in order-of-magnitude agreement with the fast component of MEH-PPV transient decay~\cite{dykstra2009conformational}, and with the ultrafast depolarization timescale~\cite{wells2008correlated}. Furthermore, in the weak-hopping regime of Figs.~2(c) and 3(c), the exchange between $|2\rangle$ and $|3\rangle$ via FRET channel emerges as a slow transfer pathway operating on the ps timescale. This is consistent with the slow component of intra-chain migration ($\sim 1$--$10$ ps)~\cite{dykstra2009conformational}.

The strongly suppressed EET observed for the 2--10 system at intermediate and weak hopping (Figs.~4(b)--(c) and 5(b)--(c)) qualitatively reflects the experimentally known asymmetry of spectral diffusion in MEH-PPV, in which exciton population preferentially migrates toward longer segments (lower-energy) and rarely transfers to chromophores with a large energy mismatch~\cite{okazaki2024generation,okazaki2023luminescence,kim2006detailed,yu2004low}.

\section{Conclusion}
\label{sec:Con} 

In this study, intra-chain EET in a conjugated polymer was analyzed using a three-level D--A model combined with HEOM-based population dynamics.
Systematic case studies were performed for two representative D--A combinations of chromophores, the 5--7 and 2--10 cases, corresponding to small and large D--A energy differences, respectively.

For the 5--7 case, where the D--A energy gap is small, strong hopping leads to population exchange showing coherent oscillation between donor and acceptor chromophores.
As the hopping strength is reduced, coherent transfer is progressively diminished, and population relaxation toward vibrationally relaxed states becomes dominant. Moreover, the FRET-mediated transition contributes on longer time scales.
By contrast, the 2--10 case is characterized by a large D--A energy gap, for which EET is strongly inhibited unless the hopping interaction is sufficiently large, and the population dynamics are governed by vibrational relaxation and localization. As a consequence, under most hopping conditions for the case of a large D--A energy gap, efficient EET between two chromophores does not occur, and the population dynamics instead converge toward thermally localized states.

For both cases, variation of $\gamma$ shows that smaller values allow coherence to persist over longer time scales, whereas larger values enhance damping and accelerate relaxation toward localized populations.

By comparing the two cases, our results suggest that EET can occur relatively easily when the two chromophores have similar site energies rather than a large energy difference. While this work considers EET between two chromophores, our model can be extended to multi-chromophore systems to mimic a segment of a polymer chain. 
Accordingly, we expect to observe stepwise EET among chromophores. Such a stepwise transfer may also contribute to understanding the ultrafast anisotropic transfer. As discussed in Section \ref{Result.C}, the timescale-level agreement with experimental reports constitutes a partial validation. A one-to-one match to spectroscopic lineshapes is, however, intrinsically limited in the reduced D--A model: the chain length is restricted and the configurational averaging too small to yield statistically meaningful spectroscopic observables so that full spectroscopic validation over multi-chromophore ensembles is left to future work.

Overall, we have established a theory in which the state energy gap, hopping constant, and S--B relaxation rate jointly determine intra-chain EET and population dynamics relevant to emission spectrum in conjugated polymers.

\section*{Acknowledgements}
This research was supported by a fellowship of the JST SPRING program, The Establishment of University Fellowships Toward the Creation of Science and Technology Innovation (Grant No. JPMJSP2110).

\end{document}